\documentclass[
  reprint,
  superscriptaddress,
  amsmath,amssymb,
  aps,
  prl,
  floatfix,
]{revtex4-1}

\usepackage{graphicx}
\usepackage{dcolumn}
\usepackage{bm}
\usepackage{hyperref}
\hypersetup{
  colorlinks=true,
  anchorcolor=blue,
  linkcolor=blue,
  citecolor=blue,
  filecolor=blue,
  urlcolor=blue
}
\usepackage{lineno}

\begin{document}

\title{Candidates for the 5$\alpha$ condensed state in ${}^{20}$Ne}

\author{S.~Adachi}
\email{adachi@ne.phys.sci.osaka-u.ac.jp}
\affiliation{Department of Physics, Osaka University,
  Machikaneyama, Toyonaka, Osaka, Japan 560-0043}%
\author{Y.~Fujikawa}%
\affiliation{Department of Physics, Kyoto University,
  Kitashirakawa-Oiwake, Sakyo, Kyoto 606-8502, Japan}%
\author{T.~Kawabata}%
\affiliation{Department of Physics, Osaka University,
  Machikaneyama, Toyonaka, Osaka, Japan 560-0043}%
\author{H.~Akimune}%
\affiliation{Department of Physics, Konan University,
  Higashinada, Kobe, Hyogo 658-8501, Japan}%
\author{T.~Doi}
\affiliation{Department of Physics, Kyoto University,
  Kitashirakawa-Oiwake, Sakyo, Kyoto 606-8502, Japan}%
\author{T.~Furuno}
\affiliation{Research Center for Nuclear Physics (RCNP),
  Osaka University, Ibaraki, Osaka 567-0047, Japan}%
\author{T.~Harada}
\affiliation{Department of Physics, Kyoto University,
  Kitashirakawa-Oiwake, Sakyo, Kyoto 606-8502, Japan}%
\author{K.~Inaba}
\affiliation{Department of Physics, Kyoto University,
  Kitashirakawa-Oiwake, Sakyo, Kyoto 606-8502, Japan}%
\author{S.~Ishida}%
\affiliation{Cyclotron and Radioisotope Center (CYRIC),
  Tohoku University, Sendai, Miyagi 980-8578, Japan}%
\author{M.~Itoh}
\affiliation{Cyclotron and Radioisotope Center (CYRIC),
  Tohoku University, Sendai, Miyagi 980-8578, Japan}%
\author{C.~Iwamoto}
\affiliation{
  RIKEN Center for Advanced Photonics, RIKEN,
  Hirosawa, Wako, Saitama 351-0198, Japan}%
\author{N.~Kobayashi}
\affiliation{Research Center for Nuclear Physics (RCNP),
  Osaka University, Ibaraki, Osaka 567-0047, Japan}%
\author{Y.~Maeda}
\affiliation{Faculty of Engineering, University of Miyazaki,
  Gaukuen-Kibanadai, Miyazaki 889-2192, Japan}%
\author{Y.~Matsuda}
\affiliation{Cyclotron and Radioisotope Center (CYRIC),
  Tohoku University, Sendai, Miyagi 980-8578, Japan}%
\author{M.~Murata}
\affiliation{Research Center for Nuclear Physics (RCNP),
  Osaka University, Ibaraki, Osaka 567-0047, Japan}%
\author{S.~Okamoto}
\affiliation{Department of Physics, Kyoto University,
  Kitashirakawa-Oiwake, Sakyo, Kyoto 606-8502, Japan}%
\author{A.~Sakaue}
\affiliation{Nishina Center for Accelerator Based Science,
  RIKEN, Hirosawa, Wako, Saitama 351-0198, Japan}%
\author{R.~Sekiya}
\affiliation{Department of Physics, Kyoto University,
  Kitashirakawa-Oiwake, Sakyo, Kyoto 606-8502, Japan}%
\author{A.~Tamii}
\affiliation{Research Center for Nuclear Physics (RCNP),
  Osaka University, Ibaraki, Osaka 567-0047, Japan}%
\author{M.~Tsumura}
\affiliation{Department of Physics, Kyoto University,
  Kitashirakawa-Oiwake, Sakyo, Kyoto 606-8502, Japan}%

\date{\today}

\begin{abstract}
  We conducted the coincidence measurement of
  $\alpha$ particles inelastically scattered from ${}^{20}$Ne
  at $0^{\circ}$ and decay charged particles
  in order to search for the alpha-particle condensed state.
  We compared the measured excitation-energy spectrum
  and decay branching ratio with the statistical-decay-model calculations,
  and found that the newly observed states
  at $E_x$ = 23.6, 21.8, and 21.2 MeV in ${}^{20}$Ne
  are strongly coupled to
  a candidate for the 4$\alpha$ condensed state in ${}^{16}$O.
  This result presents the first strong evidence
  that these states are the candidates
  for the 5$\alpha$ condensed state.
\end{abstract}

\maketitle

The alpha-cluster correlation
between two protons and two neutrons is a very important
property of atomic nuclei.
Because an alpha particle consisting of four nucleons
is tightly bound and has no excited state up to 20 MeV,
it behaves as a well-established subunit in nuclei.
It is theoretically suggested that
alpha clusters might condense into
the same lowest $0s$ orbit
in their common mean field
at low densities and temperatures
due to their bosonic nature.
This is the Bose-Einstein condensation
in the nucleon many-body systems~\cite{Tohsaki2001}.

At lower densities than the nuclear saturation density
$n_0 \approx 0.16$ ${\rm fm}^{-3}$
and low temperatures,
nuclear matter is no longer uniform and
the system minimizes its energy by forming
clusters such as deuterons, tritons, helium-3s,
and alpha particles.
The alpha particles, which are
the most tightly bound among these clusters,
are deposited in nuclear matter below a critical density
and form the alpha-particle condensate.

One of the remarkable effects
of the alpha-particle condensation
is the enhancement of the symmetry energy of nuclear matter.
The symmetry energy is conventionally
defined as the quadratic coefficient
when the internal energy per nucleon
of nuclear matter is expanded as
the Taylor series of the asymmetry parameter
$\delta = (n_n-n_p)/n_B$.
Here, $n_n$, $n_p$, and $n_B$ denote
the number densities of neutrons, protons,
and baryons, respectively.
The density and temperature dependence
of the symmetry energy is of great importance
to describe nuclear matter.
If the formation of clusters in nuclear matter
is taken into account,
the internal energies per nucleon are
considerably lowered around $\delta = 0$,
particularly at low temperatures.
Therefore, the symmetry energy, which is
the curvature of the internal energy
with respect to $\delta$,
substantially increases
at low densities below $n_B \sim 10^{-2}$
${\rm fm}^{-3}$~\cite{Typel2010,Typel2014,Zhang2019}.
The equation of state (EoS) of nuclear matter is hence
influenced by the alpha-particle condensation.
Construction of the EoS is
one of the ultimate goals in nuclear physics
not only because it is the benchmark of our understanding
about strongly interacting fermions
but also because it is required
to describe many astrophysical phenomena
such as neutron stars, supernovae,
and the nucleosynthesis in the universe.

The alpha-particle condensation is expected to manifest
as the alpha-particle condensed states (ACSs)
in finite nuclei as well as in nuclear matter.
We can study dilute nuclear matter
by examining the ACSs even though infinite nuclei
can not be formed on the earth.
The properties of the ACSs
such as energies and widths
will shed light on low-density nuclear matter.

The authors of Refs.~\cite{Tohsaki2001,Funaki2002} proposed
the Tohsaki-Horiuchi-Schuck-R\"{o}pke (THSR) wave function
to describe the $0^+$ states in ${}^8$Be,
${}^{12}$C, and ${}^{16}$O,
and theoretically suggested
the ACSs emerge
near the $2\alpha$, $3\alpha$, and $4\alpha$-decay
threshold energies, respectively.
The THSR wave function demonstrates that
these states are low-density states composed of
weakly interacting alpha particles
condensed into the lowest $0s$ orbit,
and are akin to the alpha-particle condensate in nuclear matter.
The following work~\cite{Yamada2004}
predicted that similar ACSs should exist
slightly above $k\alpha$-decay thresholds
in heavier self-conjugate $A = 4k$ nuclei
up to $k \sim 10$.
In order to establish the alpha-particle condensation
as a dilute phase of nuclear matter,
it should be examined whether the ACSs universally exist
in heavier nuclei.
However, the experimental information on
the ACSs was obtained
in limited nuclei so far.

Let us briefly describe the present situation
on the ACSs
in the self-conjugate $A = 4k$ nuclei.
The ground state of ${}^{8}$Be
and the $0_2^+$ state in ${}^{12}$C
locate near the 2$\alpha$ and 3$\alpha$-decay thresholds,
and are nicely described with the spatially developed
wave functions by the fully microscopic
alpha-cluster models~\cite{Funaki2002,Kamimura1981,Uegaki1979}
and the Green's function Monte Carlo calculation~\cite{Wiringa2000}.
These wave functions
reasonably well reproduce the energies and
inelastic form factors of these states
and are almost equivalent to the THSR wave functions
for the $2\alpha$ and $3\alpha$ condensed
states~\cite{Funaki2002,Funaki2003}.
These facts are strong evidence that
the ground state of ${}^{8}$Be and
the $0_2^+$ state in ${}^{12}$C are
the ACSs.
The exotic structures of these ACSs
highly motivate further theoretical studies
with various models such as
the antisymmetrized molecular dynamics method~\cite{Kanada-Enyo2012},
the fermionic molecular dynamics method~\cite{Chernykh2007},
the chiral effective field theory~\cite{Epelbaum2012}, and
the algebraic cluster model~\cite{Marin-Lambarri2014,Bijker2020}.
The ACSs are discussed also in non-self-conjugate
nuclei, {\it{e.g.}} in Refs.~\cite{Yamada2015,Inaba2020}.

The ACS in ${}^{16}$O was
theoretically predicted to be the $0_6^+$ state
with the width of 140 keV~\cite{Funaki2008,Yamada2012,Funaki2018}.
The excited state at $E_x = 15.097 \pm 0.005$ MeV
with the width of $166 \pm 30$ keV~\cite{Tilley1993}
is proposed to be the corresponding state.
Although this state was observed
in many reactions~\cite{Tilley1993},
it was pointed out in Ref.~\cite{Li2017}
that this state might be contaminated
with a previously unidentified resonance
which does not exhibit $0^+$ character.
Recently it was found that
this state decays into the
${}^{8}{\rm Be}(0_1^+)+{}^{8}{\rm Be}(0_1^+)$ or
${}^{12}{\rm C}(0_2^+)+\alpha$ channel
with the almost same probabilities~\cite{Barbui2018}.
The $0_6^+$ state in ${}^{16}$O is, therefore,
a strong candidate of the ACS.

For ${}^{20}$Ne and heavier nuclei,
no known states are assigned
to the ACSs.
Only a tentative candidate for
the $5\alpha$ condensed state
was experimentally proposed.
The several $0^+$ states observed
in the ${}^{22}{\rm Ne}(p,t){}^{20}{\rm Ne}$ reaction
were examined with the shell-model
calculation~\cite{Swartz2015}.
It was found that one of the $0^+$ states at
$E_x = 22.5$ MeV was not described by the shell model,
and its excitation energy is close to
$E_x =$ 21.14 MeV where the $5\alpha$ condensed state
is predicted in Ref.~\cite{Yamada2004}.
However,
the excitation energy is not conclusive evidence
for the ACS
because many $0^+$ states exist
around the expected excitation energy.
Further information is necessary
to identify the ACS.
It should be noted that
the decay property of the excited state
provides additional information.
Because all the alpha particles in the ACSs
occupy the same lowest $0s$ orbit,
the wave functions of the alpha particles
in the ACSs of different nuclei
are similar and
the overlap between them should be large.
Therefore, ACSs
should prefer to decay via ACSs in lighter nuclei
by emitting alpha particles.
Because energy differences between ACSs
are smaller than a few MeV~\cite{Yamada2004},
emission of low-energy $\alpha$ particles
from $0^+$ states near $k\alpha$-decay thresholds
can be a clue to identify ACSs.

The ACS might be dynamically accessed via
heavy-ion collision and fusion
reaction~\cite{Vadas2015,Schuetrumpf2017}.
It is worthy to mention the recent measurement
of the ${}^{12}{\rm C}({}^{16}{\rm O},{}^{28}{\rm Si}^{*})$
reaction~\cite{Bishop2019}.
Decay alpha particles were comprehensively detected
to obtain the invariant-mass spectra of 
${}^{16}\rm O$, ${}^{20}\rm Ne$, and ${}^{24}\rm Mg$,
but no ACSs were found.
The authors claimed that the Coulomb barrier
inherently suppresses low-energy-particle decays and
obscures the signature of the ACSs.
However, some of the authors
previously pointed out that the Coulomb barrier
in the ACSs should be suppressed due to
their dilute nature~\cite{Kokalova2006}.
One plausible explanation for this contradictory situation is
that the ACSs were hidden by a lot of backgrounds
from various high-spin states in Ref.~\cite{Bishop2019}
because large angular momenta were brought to the system
in the heavy-ion collision.
Since ACSs have the spin and parity of $0^+$ and isospin of 0,
nuclear reactions which selectively excite
isoscalar $0^+$ states should be employed to populate ACSs.
One of the best reactions to populate
isoscalar $0^+$ states is inelastic alpha scattering
at forward angles~\cite{Harakeh2001}.
Because both the spin and isospin
of the alpha particle are 0,
the inelastic alpha scattering off
self-conjugate $A = 4k$ nuclei selectively
excites isoscalar natural-parity states.
In addition, the cross sections
for the $0^+$ states have
their maximum at $0^{\circ}$.

In the present work, we carried out the coincidence measurement
of alpha particles inelastically scattered from ${}^{20}$Ne
at $0^{\circ}$ and decay charged particles emitted
from excited states in order to
search for the 5$\alpha$ condensed state
in ${}^{20}$Ne.

The experiment was conducted
at the Research Center for Nuclear Physics (RCNP),
Osaka University.
The experimental setup was almost the same with Ref.~\cite{Itoh2003}
except for a ${}^{20}$Ne gas target and a Si telescope array.

A 386-MeV ${}^4{\rm He}^{2+}$ beam
with an intensity of about 10 nA
was transported to an isotopically
enriched ${}^{20}$Ne gas target.
The ${}^{20}$Ne gas with the enrichment of 99.95\%
was filled in the target cell at 14.1 kPa and room temperature.
The thickness of the target cell
along the beam axis was 8.0 mm, which corresponds to
the mass thickness of 89.6 $\mu$g/cm${}^2$
of the ${}^{20}$Ne gas.
The entrance and exit windows of the target cell
were sealed with 100 nm-thick
silicon nitride (SiN${}_{x}$) membranes.
The square membranes with the size
of 10 mm $\times$ 10 mm were glued
on the target cell.

\begin{figure}[t]
  \centering
  \includegraphics[width=8.5cm]{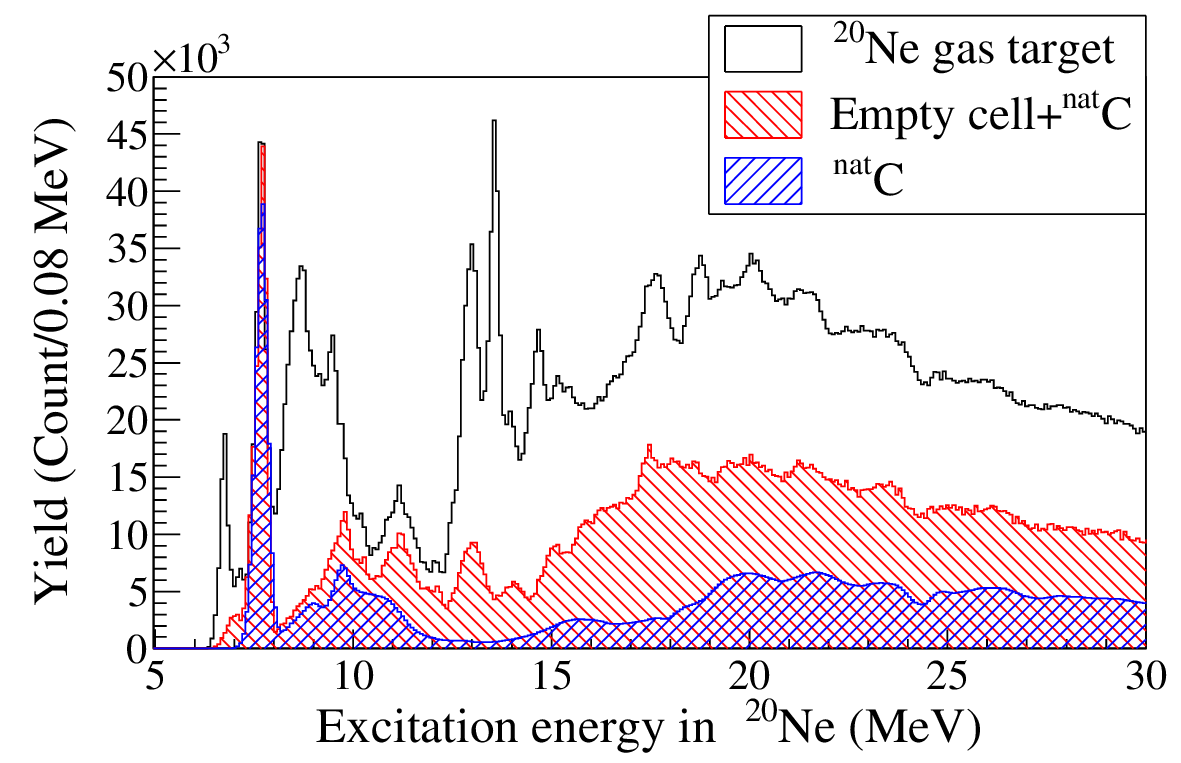}
  \caption{
    Excitation-energy spectra for
    the ${}^{20}$Ne gas target (black open),
    the empty cell with ${}^{\rm nat}$C (red hatched),
    and ${}^{\rm nat}$C foil (blue hatched).
    The spectra of the empty cell with ${}^{\rm nat}$C
    and the ${}^{\rm nat}$C foil
    are scaled for the subtraction.
  }
  \label{fig:Subtraction}
\end{figure}

Alpha particles inelastically scattered
from the target were analyzed
using the magnetic spectrometer
Grand Raiden \cite{Fujiwara1999}.
The backgrounds from the SiN${}_{x}$ membranes
were subtracted
by using the measurement with the empty cell.
We found that oil mists from vacuum pumps deposited on
the SiN${}_{x}$ membrane and
caused ${}^{\rm nat}$C background.
The ${}^{\rm nat}$C background was
successfully subtracted by using the measurement
with ${}^{\rm nat}$C foil.
Figure~\ref{fig:Subtraction} shows
the excitation-energy spectra for
the ${}^{20}$Ne gas target,
the empty cell with ${}^{\rm nat}$C,
and the ${}^{\rm nat}$C foil.

In order to detect decay particles,
we installed a Si telescope array
around the target
covering 0.42 sr (3.4\% of 4$\pi$)
at $\theta_{\rm lab} =$ 106.4--163.0${}^{\circ}$.
The Si telescope array consisted of six segments,
which were placed 170 mm away from the target.
Each segment had three layers of Si detectors,
but only the first and second layers were used
in the present analysis.
The thicknesses of the first and second layers were
65 $\mu$m and 500 $\mu$m respectively,
in order from the target side,
and their dimensions were 50 mm $\times$ 50 mm.
The first layers were divided into 8 strips and
the second layers were read as a single pad.
The first Si detectors had a dead layer
with a thickness of about 1.2 $\mu$m on the rear side.
The detection threshold of decay particles
was 0.53 MeV.

Particle identification (PID) for decay particles
that penetrated the first layer
was performed by using the correlation between
the energy loss in the first layer
and the total energy
($\varDelta E$--$E$ method). This $\varDelta E$--$E$ method
succeeded in separating decay alpha particles
from hydrogen isotopes with an accuracy of almost 100\%,
and the $\varDelta E$--$E$ method could be applied
to reject protons at higher particle energies than 2.45 MeV.
On the other hand,
the time of flight from the target to the Si detector
and the total energy were used
for low-energy particles which stopped at the first layer
(TOF method).
The time of flight was obtained from
the time difference between the radio-frequency signals
from the cyclotron and the timing signals of the Si detectors.
As the kinetic energy of decay particles decreased,
the resolution of the time of flight got worse.
Therefore, it was required to narrow the gate width
for the time of flight
to prevent the contamination due to hydrogen isotopes.
The PID accuracy and efficiency were determined
as a function of the particle energy
by analyzing the time-of-fight spectra
at different energies.
Typically, for the decay particles
with the kinetic energy of 1 MeV,
the PID efficiency for alpha particles was 69\% and
the 12\% of detected hydrogen isotopes was misidentified
as alpha particles.

\begin{figure}[t]
  \includegraphics[width=8.5cm]{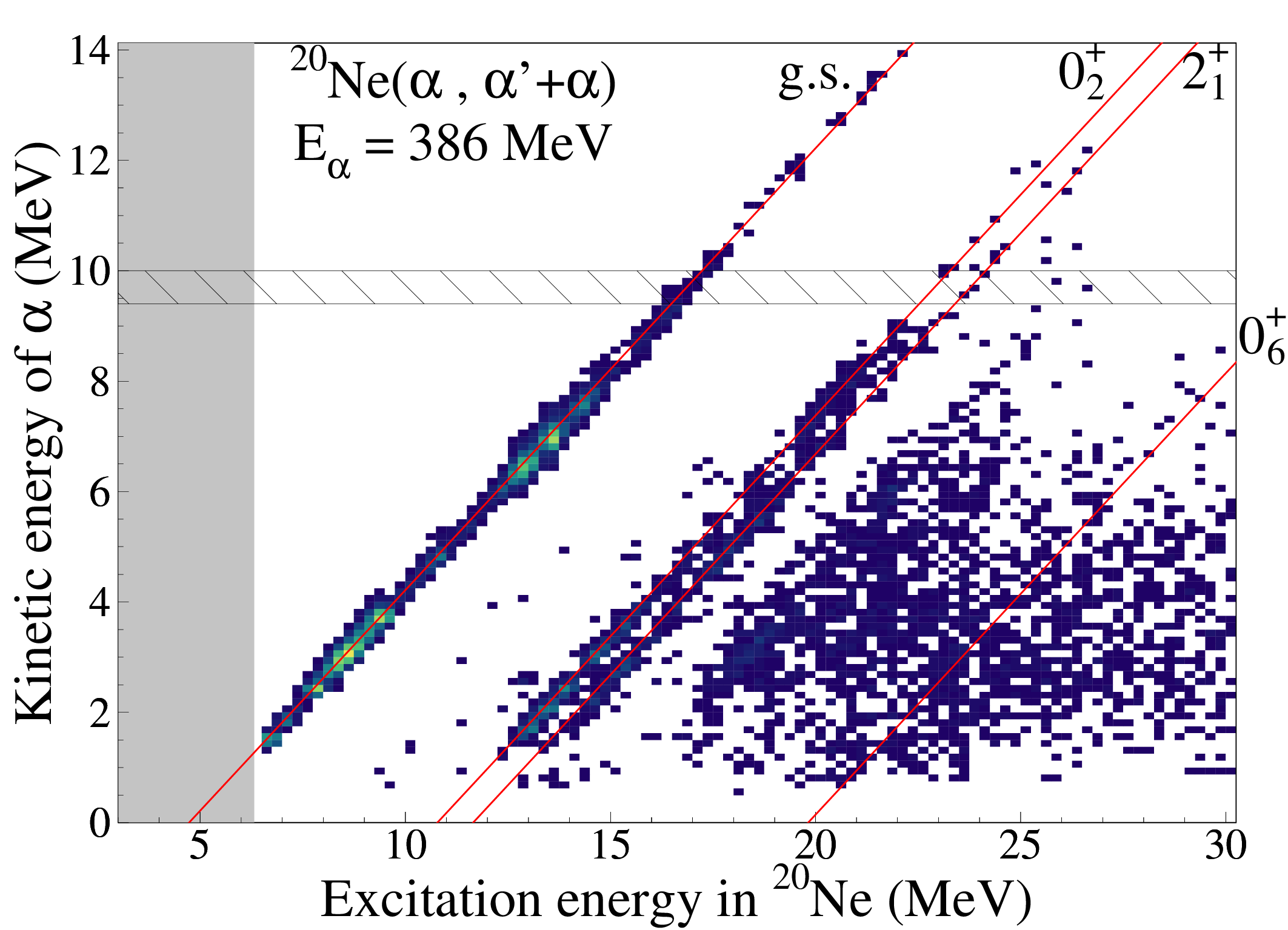}
  \caption{Correlation between the kinetic energy
    of decay alpha particles $(K_{\alpha})$
    and the excitation energies
    of ${}^{20}$Ne.
    The red solid lines indicate the calculated
    correlation in the $\alpha$-decay events
    into the ground, $0_2^+$, $2_1^+$, and $0_6^+$ states
    in ${}^{16}$O.
    The alpha particles with $K_{\alpha} < 9.7$ MeV
    stopped at the first layer of the Si detectors.
    $K_{\alpha}$ in the hatched area are accompanied
    by uncertainties due to the dead layer
    of the Si detectors.
    The shaded area represents the excitation-energy region
    out of the momentum acceptance of the Grand Raiden spectrometer.
  }
  \label{fig:CorrelationAlpha}
\end{figure}

Figure~\ref{fig:CorrelationAlpha} shows
the correlation between
the kinetic energies of decay alpha particles $(K_{\alpha})$
and the excitation energies of
${}^{20}$Ne $[E_x({}^{20}{\rm Ne})]$.
Linear loci corresponding to the $\alpha$-decay events into
the ground, $0_2^+$, and $2_1^+$ states in ${}^{16}$O
are clearly seen.
In the two-body decay of ${}^{20}$Ne into
the ${}^{16}{\rm O} + \alpha$ channel,
$K_{\alpha}$ correlates with $E_x({}^{20}\rm{Ne})$ via
the excitation energy of ${}^{16}$O
$[E_x({}^{16}\rm O)]$ as
\begin{equation*}
  K_{\alpha} =
  \dfrac{m_{{}^{16}{\rm O}}}{m_{{}^{16}{\rm O}}+m_{\alpha}}
  \left[ E_x({}^{20}{\rm Ne})
    - E_{\rm th}({}^{16}{\rm O}+\alpha) - E_x({}^{16}\rm{O})
  \right].
\end{equation*}
$E_{\rm th}({}^{16}{\rm O}+\alpha)$ is
the threshold energy for the ${}^{16}{\rm O}+\alpha$ decay
in ${}^{20}$Ne, and
$m_{{}^{16}{\rm O}(\alpha)}$ is the rest mass
of ${}^{16}$O (the $\alpha$ particle).

\begin{figure}[t]
  \includegraphics[width=8.5cm]{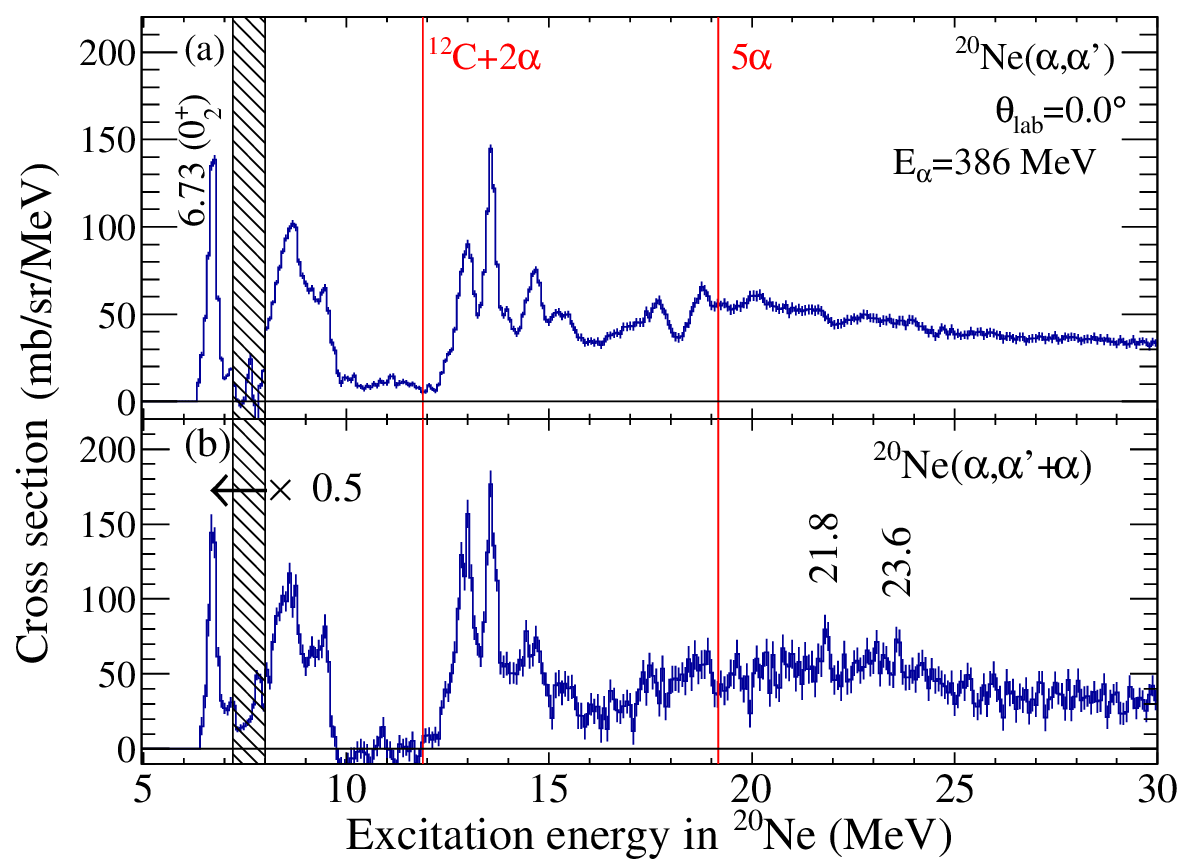}
  \caption{Excitation-energy spectra of
    the ${}^{20}{\rm Ne}(\alpha,\alpha')$ reaction at $0^{\circ}$
    for (a) the singles events and (b) the coincidence events.
    The error bars in the lower panel include the systematic errors
    in the PID analysis as well as the statistical errors.
    The spectra at $E_x < 8.0$ MeV are
    downscaled by a factor of 0.5.
    The hatched area indicates the excitation-energy region
    where a strong peak due to the $0_2^+$ state in the ${}^{12}$C
    contaminants caused large errors in the background subtraction.
    The vertical lines at $E_x =$ 11.9 and 19.2 MeV represent
    the ${}^{12}{\rm C}+2\alpha$ and $5\alpha$ decay thresholds.
  }
  \label{fig:ExSpectra}
\end{figure}

Figure~\ref{fig:ExSpectra}(a) shows
the excitation-energy spectrum of
the ${}^{20}{\rm Ne}(\alpha,\alpha')$ reaction
at $0^{\circ}$ for the singles events
obtained by measuring inelastically scattered alpha particles only.
No clear peaks are observed around 
a few MeV above the $5\alpha$ decay threshold
where the ACS is expected.
Figure~\ref{fig:ExSpectra}(b) shows the excitation-energy spectrum
for the coincidence events in which one alpha particle was detected
by the Si telescope array.
The obtained yield was corrected
with the PID efficiency for alpha particle, and
converted to the cross section
on the basis of the fact that the $\alpha$-decay probability of
the $0_2^+$ state at $E_x = 6.73$ MeV in ${}^{20}$Ne is
almost 100\%~\cite{Tilley1998}
The error bars of the spectrum in Fig.~\ref{fig:ExSpectra}(b)
are calculated by quadratically adding
the statistical errors and the systematic errors in the PID analysis.
Note that the cross sections
for the ${}^{20}{\rm Ne}(\alpha,\alpha'+\alpha)$ reaction
in Fig.~\ref{fig:ExSpectra}(b) can be apparently larger than
those for the ${}^{20}{\rm Ne}(\alpha,\alpha')$ reaction
in Fig.~\ref{fig:ExSpectra}(a)
above the ${}^{12}{\rm C}+2\alpha$ threshold
because more than one alpha particle is
allowed to be emitted.
It is remarkable that the two narrow peaks are
observed at $E_x =$ 23.6 and 21.8 MeV
above the $5\alpha$ decay threshold,
where no peaks are seen in Fig.~\ref{fig:ExSpectra}(a)
although their statistical significance is not large.

\begin{figure}[t]
  \centering
  \includegraphics[width=8.5cm]{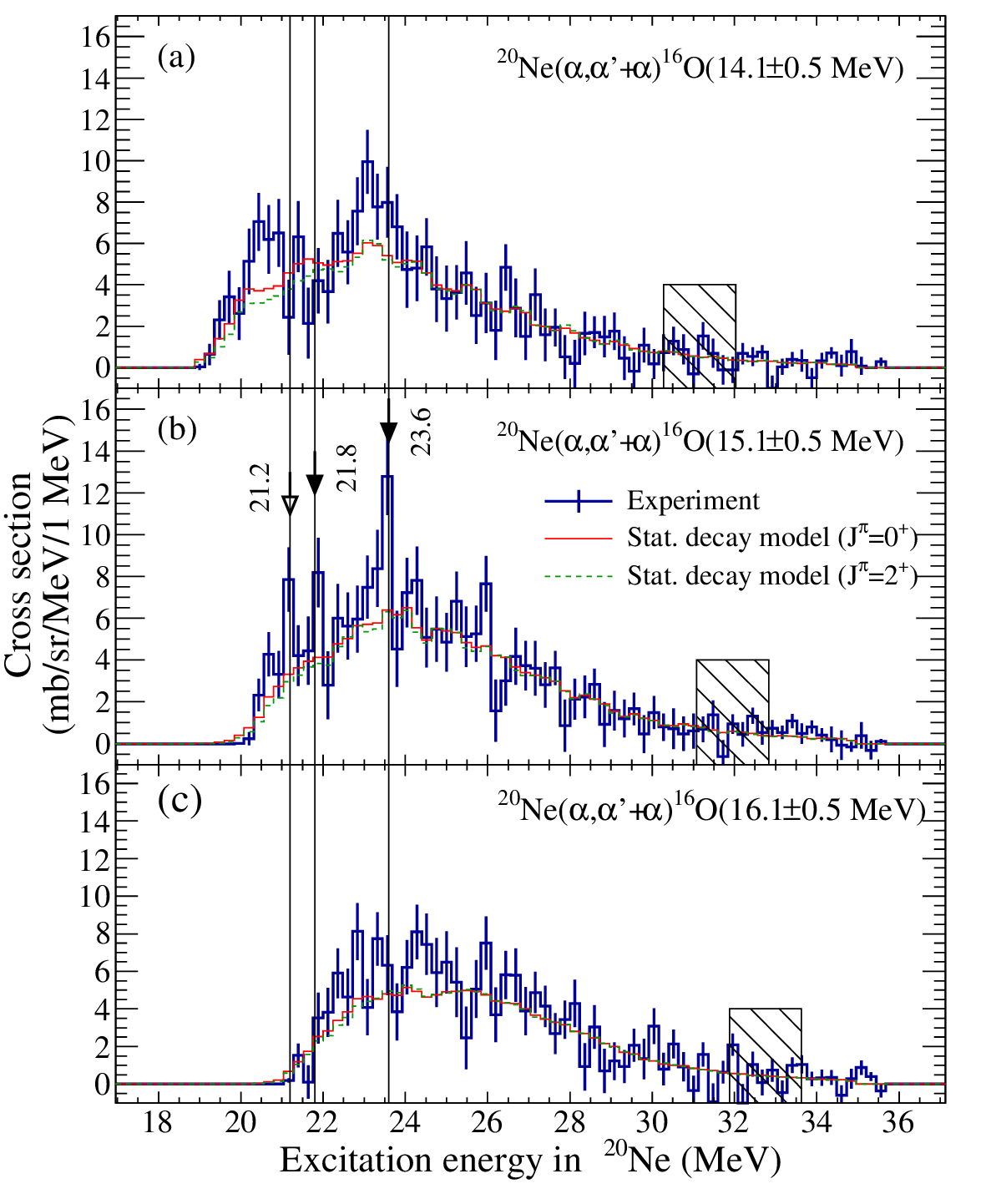}
  \caption{
    Excitation-energy spectra
    same as Fig.~\ref{fig:ExSpectra}(b)
    but when $E_x({}^{16}\rm O)$ is
    (a) $14.1 \pm 0.5$ MeV,
    (b) $15.1 \pm 0.5$ MeV, and (c) $16.1 \pm 0.5$ MeV.
    The statistical-decay-model calculations
    for the $0^+$ and $2^+$ states in ${}^{20}$Ne
    are shown together with thin red solid and green dashed lines.
    The vertical lines at 21.2, 21.8, and 23.6 MeV are drawn
    to guide the eyes.
    The dead layers of the Si detectors caused
    the uncertainty on the kinetic energies
    of alpha particles in the hatched area
    as for Fig.~\ref{fig:CorrelationAlpha}.
  }
  \label{fig:Cascade1}
\end{figure}

In order to search for the $5\alpha$ condensed state,
we focused on the $\alpha$-decay events
into the $0_6^+$ state in ${}^{16}$O
because this state is
a strong candidate for the 4$\alpha$ condensed state.
We reconstructed $E_x({}^{16}\rm O)$
from $K_{\alpha}$ and $E_x({}^{20}\rm Ne)$
assuming the two-body
${}^{20}{\rm Ne}^{*}\rightarrow{}^{16}{\rm O}+\alpha$ decay
and selected the events with
$E_x({}^{16}\rm O) = 15.1\pm0.5$ MeV
to obtain the excitation-energy spectrum of
${}^{20}$Ne in the
${}^{20}\mathrm{Ne}(\alpha,\alpha'+\alpha){}^{16}\mathrm{O}(0_6^+)$
reaction
as shown in Fig.~\ref{fig:Cascade1}(b).
The same normalization factor with Fig.~\ref{fig:ExSpectra}(b)
was used to convert the yield to the cross section.
The similar spectra for the events in the
1.0-MeV lower and higher $E_x({}^{16}\rm O)$ ranges
were also shown in Figs.~\ref{fig:Cascade1}(a) and (c)
for comparison.
The error bars in Fig.~\ref{fig:Cascade1} include the
statistical errors and the systematic errors
in the PID analysis as in Fig.~\ref{fig:ExSpectra}(b).
For example, the cross section
at $E_x({}^{20}\mathrm{Ne}) = 21.2$ MeV
in Fig.~\ref{fig:Cascade1}(b)
is $7.9\pm1.3\,(\mathrm{stat.})\pm0.8\,(\mathrm{sys.})=7.9\pm1.5$
mb/sr/MeV/1 MeV.

The excitation-energy spectra in Fig.~\ref{fig:Cascade1}
were compared with the theoretical spectra calculated
by the statistical-decay model.
This model takes into account the spins, parities, isospins,
energies, and level densities of
the mother and its descendant nuclei
as well as transmission coefficients for decay particles.
Decay branching ratios of excited states in ${}^{20}$Ne
into particle- ($\alpha$, $p$, and $n$)
and $\gamma$-decay channels were
calculated with the computer code CASCADE \cite{Puhlhofer1977}
assuming that the initial excited state is the isoscalar state
with its spin and parity of $0^+$, $1^-$, or $2^+$.
Using the theoretical branching ratios,
we performed the Monte Carlo calculation
to simulate the decay processes
of the excited states in ${}^{20}$Ne.
The decay processes were traced until all of
the descendant nuclei settled in their ground states
under the assumption that decay particles were emitted
isotropically in the rest frame of the decaying nuclei.
The simulated decay events were analyzed in the same manner
with the experimental data
after the energies of decay particles were randomly varied
according to the experimental resolution of 0.40 MeV
at the full width at half maximum.
The theoretical spectra for the $0^+$, $1^-$, and $2^+$ states were
almost the same,
and
the cross section for the $1^-$ state
in the inelastic alpha scattering is generally
smaller than those for the $0^+$ and $2^+$
states at $0^{\circ}$.
Therefore, only the theoretical spectra for
the $0^+$ and $2^+$ states are presented
in Fig.~\ref{fig:Cascade1}.
The theoretical spectra in Figs.~\ref{fig:Cascade1}(a), (b), and (c)
were multiplied by a factor of 0.74.
This factor is determined in Fig.~\ref{fig:Cascade1}(b)
to fit the theoretical spectrum for the $0^+$ state
to the experimental spectrum
at $E_x = 27$--31 MeV
where no structures were observed in the experimental spectrum.
In contrast to the good agreement between
the theoretical calculations and the experiment
above $E_x = 26$ MeV in Fig.~\ref{fig:Cascade1},
the two peaks at $E_x = 23.6$ and 21.8 MeV,
which were observed in Fig.~\ref{fig:ExSpectra}(b),
were visible in Fig.~\ref{fig:Cascade1}(b)
as indicated by the solid arrows
but not in Figs.~\ref{fig:Cascade1}(a) and (c).
It demonstrates that
these two states
strongly couple to the $0_6^+$ state in ${}^{16}$O.
The experimental cross section
in Fig.~\ref{fig:Cascade1}(b)
also exceeds the calculations
at $E_x$ = 21.2 MeV as indicated by the open arrow,
where no clear peak structures are seen
in Fig.~\ref{fig:ExSpectra}(b).
This peak is not observed
in Figs.~\ref{fig:Cascade1}(b) and (c).
The 21.2-MeV state is closer
to the $5\alpha$ threshold
than the two states at $E_x =$ 23.6 and 21.8 MeV.
The similar excess is also observed
in $E_x = 24$--26 MeV
although the statistical uncertainty is large.

\begin{figure}[t]
  \centering
  \includegraphics[width=8.5cm]{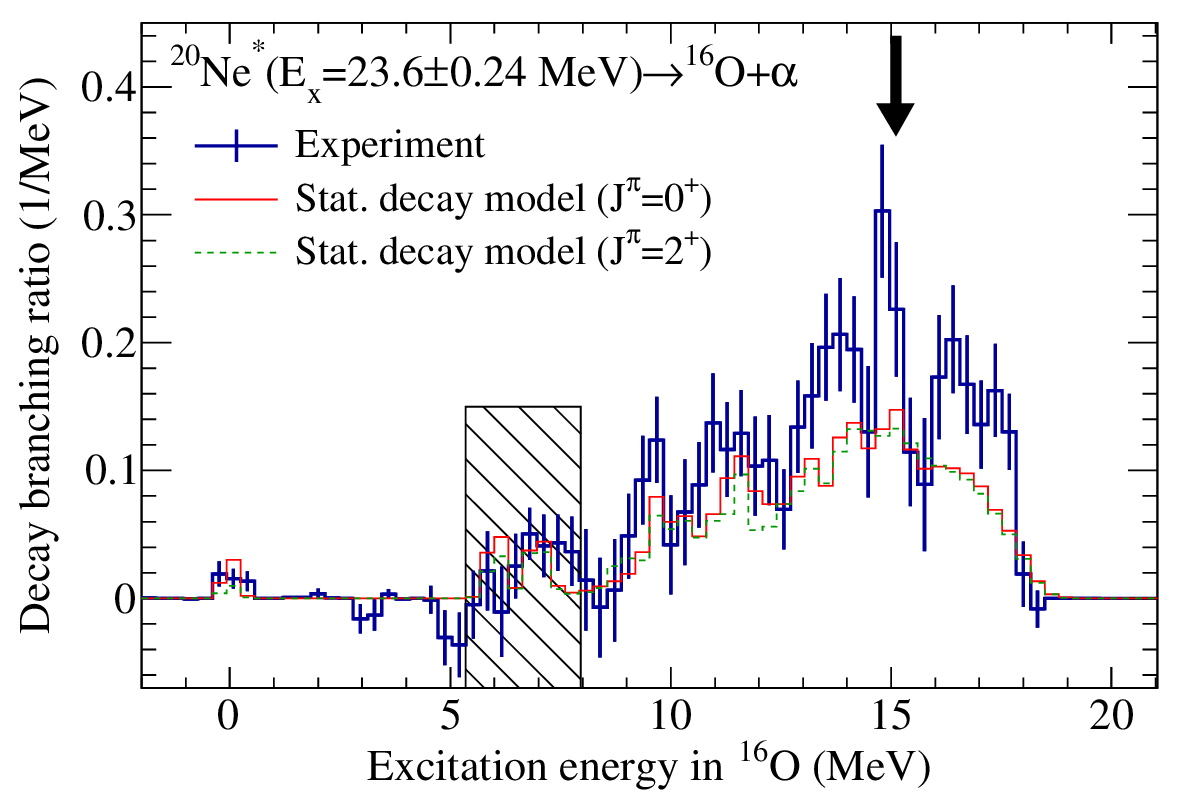}
  \caption{Decay branching ratio of the excited state
    in ${}^{20}$Ne at $E_x = 23.6 \pm 0.24$ MeV populated
    by the inelastic alpha scattering at $0^{\circ}$
    (thick blue solid lines with error bars) compared with
    the statistical-decay-model calculations
    from the isoscalar $0^+$ and $2^+$ states in ${}^{20}$Ne
    (thin red solid and green dashed lines).
    The dead layers of the Si detectors caused
    the uncertainty on the kinetic energies
    of alpha particles in the hatched area
    as for Fig.~\ref{fig:CorrelationAlpha}.
  }
  \label{fig:Cascade2}
\end{figure}

In order to examine the decay properties of
the new state at $E_x$ = 23.6 MeV in ${}^{20}$Ne,
the $\alpha$-decay events
at $E_x = 23.6 \pm 0.24$ MeV were selected,
and the decay branching ratio into ${}^{16}$O was
obtained assuming the two-body decay
as shown in Fig.~\ref{fig:Cascade2}.
The theoretical branching ratios of
the isoscalar $0^+$ and $2^+$ states
at the same excitation energy ($E_x = 23.6 \pm 0.24$ MeV)
were obtained from the simulated decay events
using the same normalization factor as in Fig.~\ref{fig:Cascade1},
and compared with the experiment.
The theoretical branching ratios do not change much
depending on the spins of the initial excited states.
Several prominent peaks were observed on the continuous
spectrum calculated by the statistical-decay model.
It should be noted that the location of the strongest peak
around $E_x({}^{16}{\rm O})$ = 15 MeV
agrees with that of the $0_6^+$ state in ${}^{16}$O
at $E_x({}^{16}\rm O)$ = 15.1 MeV
indicated by the vertical arrow.

Figures~\ref{fig:Cascade1} and \ref{fig:Cascade2} show that
the newly found excited state at $E_x =$ 23.6 MeV
is strongly coupled to the $0_6^+$ state in ${}^{16}$O,
which is a candidate for the $4\alpha$ condensed state.
This 23.6-MeV state is a candidate for the $5\alpha$ condensed state.
However, the measured excitation energy is considerably higher than
the theoretical value of $E_x =$ 21.14 MeV~\cite{Yamada2004}.
This might suggest another interpretation that
either of the low-energy states
at $E_x =$ 21.8 or 21.2 MeV corresponds to
the $5\alpha$ condensed state, and
the 23.6-MeV state is akin to the $5\alpha$ condensed state
like the $2_2^+$ state in ${}^{12}$C,
which is an excited state of the relative motion
of alpha clusters in the $3\alpha$ condensed
state~\cite{Uegaki1979,Kamimura1981,Descouvemont1987,Funaki2015}.

The high-lying structures around $E_x =$ 24--26 MeV
in Fig.~\ref{fig:Cascade1}(b)
might be analogous to the $0^+$ states
above the 4$\alpha$ condensed state
in ${}^{16}$O, which are coupled
to the $\alpha+{}^{12}\mathrm{C}(0_2^+)$
channel as predicted in Ref.~\cite{Funaki2012}.
In another interpretation, these structures
might be due to the fragmentation of alpha-cluster states
which is discussed in medium-mass
nuclei~\cite{Norrby2011,Bailey2019}.

In order to clarify the correspondence between these new states
and the $5\alpha$ condensed state,
it is necessary to determine their spins and parities.
Regrettably, the spins and parities of the new states
could not be assigned in the present work.
However, it is worthy to mention that
the states at 21.2 and 21.8 MeV are very close to
the tentative $0^+$ states at 21.16 and 21.80 MeV
reported in Ref.~\cite{Swartz2015}.
This fact also supports their candidacy
for the 5$\alpha$ condensed state.

In summary,
we conducted the coincidence measurement of
alpha particles inelastically scattered from ${}^{20}$Ne
at $0^{\circ}$ and decay charged particles from excited states
in order to search for the $5\alpha$ condensed state
in ${}^{20}$Ne.
Comparing the measured excitation-energy spectra and
decay branching ratio
with the statistical-decay-model calculation,
we found that the newly observed states
at $E_x$ = 23.6, 21.8, and 21.2 MeV
in ${}^{20}$Ne are strongly coupled
to the $0_6^+$ state in ${}^{16}$O.
This result presents the first strong evidence
that these states are the candidates
for the $5\alpha$ condensed state in ${}^{20}$Ne
because the $0_6^+$ state in ${}^{16}$O
is a strong candidate for the $4\alpha$ condensed state.
However, their spins and parities are still ambiguous.
An additional measurement to determine
their spins and parities is strongly desired.

The authors acknowledge the RCNP cyclotron crews
for providing a high-quality beam
for background-free measurements at $0^{\circ}$.
This work was performed under
the RCNP E402 program, and
partly supported by JSPS KAKENHI
Grants No. JP20K14490, JP19J20784, and JP19H05153.

\nocite{*}


\bibliography{arxiv5.0}

\end{document}